\documentclass[graybox]{svmult}

\usepackage{type1cm}        
\usepackage{makeidx}         
\usepackage{graphicx}        
\usepackage{multicol}        
\usepackage[bottom]{footmisc}

\usepackage{newtxtext}       %
\usepackage{newtxmath}       

\usepackage[utf8]{inputenc}
\usepackage{color}
\usepackage{url}
\usepackage{graphicx}
\usepackage[numbers]{natbib}
\usepackage{multirow}
\usepackage{hhline}
\usepackage{booktabs}
\usepackage{colortbl}

\begin{document}

\title*{Wrist movement classification for adaptive mobile phone based rehabilitation of children with motor skill impairments}
\titlerunning{Wrist movement classification for rehabilitation of children}

\author{Kayleigh Schoorl and Tamara Pinos Cisneros and Albert Ali Salah and Ben Schouten}
\authorrunning{K. Schoorl et al.}

\institute{
Kayleigh Schoorl \at Utrecht University, Utrecht, the Netherlands,
\email{kayleighschoorl@gmail.com}
\and Albert Ali Salah \at Utrecht University, Utrecht, the Netherlands 
\email{a.a.salah@uu.nl}
\and 
Tamara Pinos Cisneros \at University of Twente and Amsterdam University of Applied Sciences, the Netherlands
\email{t.v.pinos.cisneros@hva.nl}
\and
Ben Schouten \at Eindhoven University of Technology, the Netherlands
\email{bschouten@tue.nl}
}


\maketitle
\abstract{
Rehabilitation exercises performed by children with cerebral palsy are tedious and repetitive. To make them more engaging, we propose to use an exergame approach, where an adaptive application can help the child remain stimulated and interested during exercises. In this paper, we describe how the mobile phone sensors can be used to classify wrist movements of the user during the rehabilitation exercises to detect if the user is performing the correct exercise and illustrate the use of our approach in an actual mobile phone application. We also show how an adaptive difficulty system was added to the application to allow the system to adjust to the user. We present experimental results from a pilot with healthy subjects that were constrained to simulate restricted wrist movements, as well as from tests with a target group of children with cerebral palsy. Our results show that wrist movement classification is successfully achieved and results in improved interactions. 
}


\section{Introduction}
Children with cerebral palsy often need to perform repetitive movements during these therapy sessions. It can be a difficult task to keep them motivated to continue doing these exercises regularly, especially when these need to be performed at home in between sessions with their therapists. When they are intrinsically motivated, they participate in therapy for their own satisfaction and are more likely to improve and keep up with their therapy sessions and exercises~\cite{ryan2000intrinsic}. Moreover, children with cerebral palsy might often encounter frustrating moments and failure during their therapy, which underlines the need for keeping therapy as motivational and engaging as possible to minimize frustrations and uphold motivation~\cite{fikar2018use}. It is therefore important for a therapist to accommodate an environment that both enhances self-directed engagement in therapeutic exercises and also provides intrinsic motivation to perform the therapy exercises at home.

Since the exercises recommended for cerebral palsy are often not inherently experienced as being enjoyable, applied games can be a powerful tool to motivate children to gain more intrinsic motivation, and can contribute to a higher degree of personalization and adaptivity to accommodate their individual needs~\cite{salah2014playful, cisneros2020hand}. Digital developments make different types of learning incorporated into video games possible, allowing for games that are both engaging to play and can help children to learn~\cite{shaffer2006computer}. They can help decrease the negative aspects of therapy, such as frustration and boredom, and increase the positive aspects, such as happiness and enjoyment, and can therefore be a suitable tool for providing intrinsic motivation for children to perform their therapeutic exercises.

The hand and wrist rehabilitation of cerebral palsy patients focuses on a type of movement called ``dorsiflexion". The dorsiflexion of the wrist and hand involves bending the hand upward towards the top of the forearm; the downward bending movement towards the bottom of the forearm is called ``palmar flexion". The main exercise for the patients is shaking the hand up and down, moving between dorsiflexion and palmar flexion repeatedly. The mobile rehabilitation application requires such movements for successful completion of its tasks.

In this paper, we describe how a mobile application that has been developed for facilitating hand therapy~\cite{alberts2022designing} can be extended to become adaptive by classifying hand and wrist dorsiflexion movements of its users on the fly. For this purpose, we test the suitability of a small number of classifiers, based on the sensors on the phone. We test and refine our approach first on a pilot study with healthy subjects, whose wrists we constrain in several ways to simulate patient conditions. Our original target group for this application is children from ages 7 to 12, who have hemiparesis as a result of cerebral palsy and are within level I or II on the Manual Ability Classification Scale or MACS~\cite{eliasson2006manual}. After the pilot, we also conduct an experiment with eight children with cerebral palsy, and present our results. 

\begin{figure}[thb]
  \centering
  \includegraphics[height=6cm]{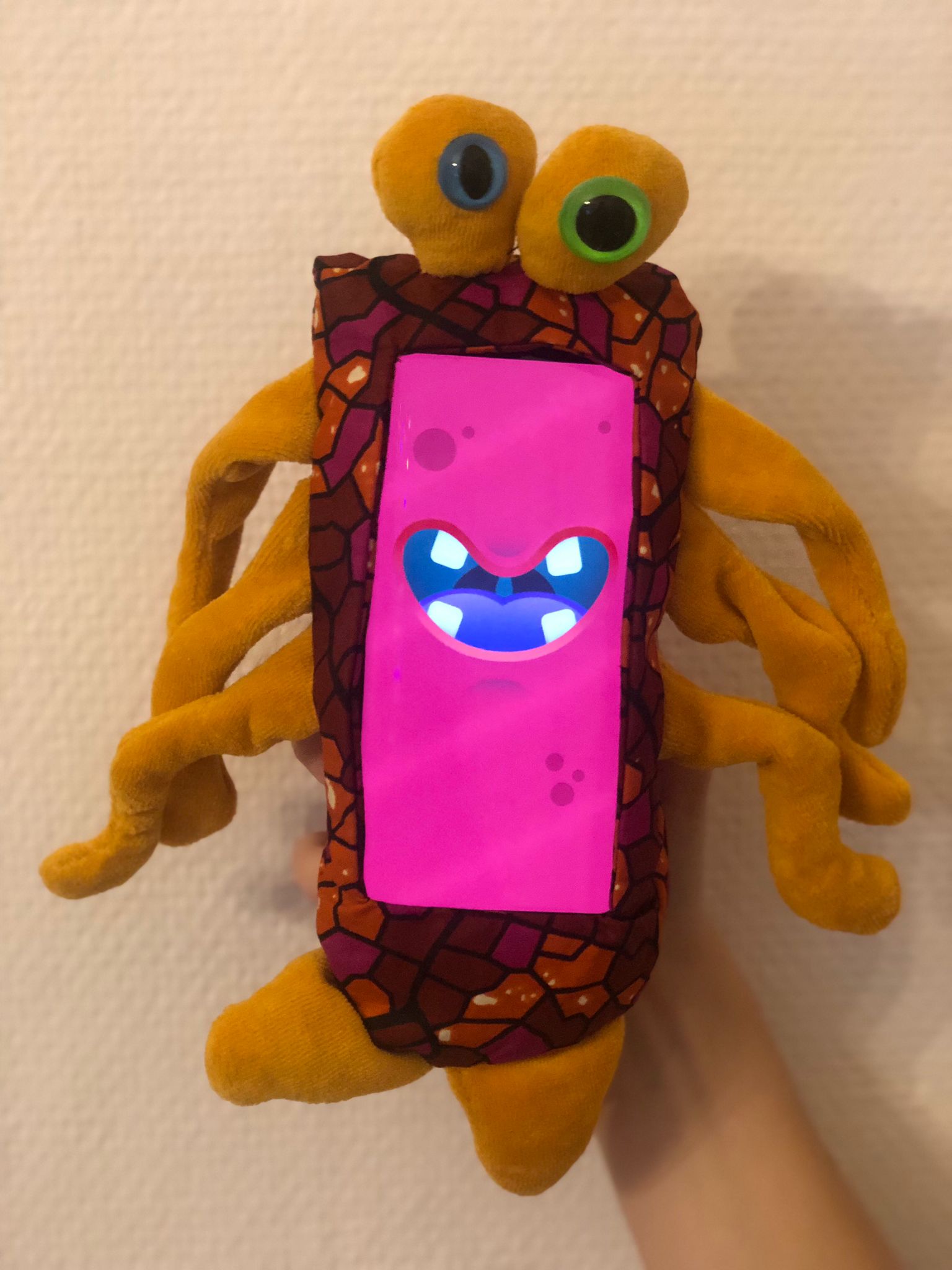}
  \includegraphics[height=6cm]{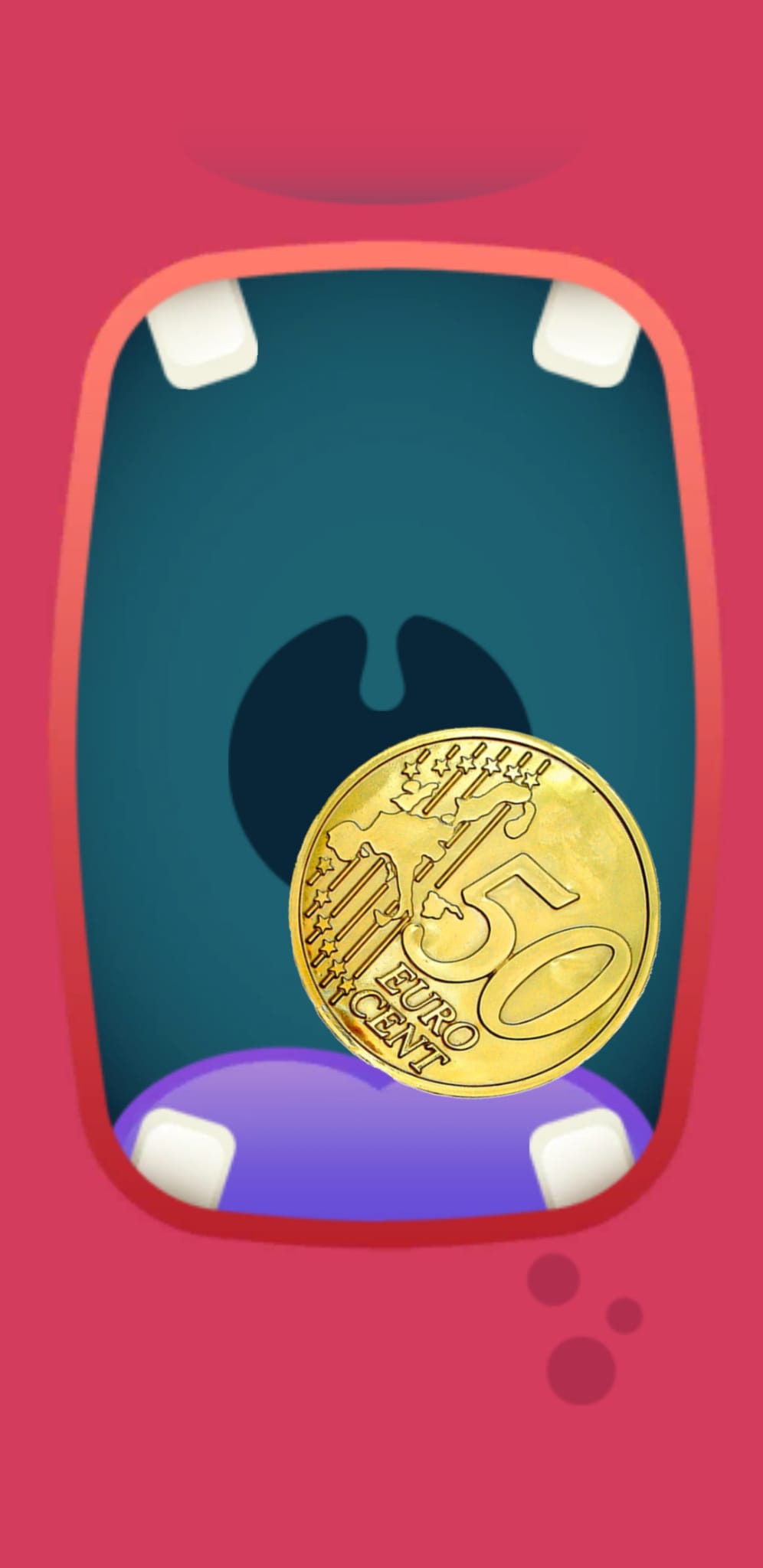}
  \caption{The monster case used for the Magic Monster application and a screenshot from the mobile app.}
  \label{fig:smarttoy}
\end{figure}

We briefly introduce the mobile phone based exergame application here, which serves as our experimental platform. The phone is inserted in a soft, animal-shaped sheath, and the phone display becomes the face of the animal, which is capable of expressing emotions, as can be seen in Figure~\ref{fig:smarttoy}. In the game, the animal is an assistant for performing magic tricks and comes with a separate booklet in which the magic tricks are explained. The performing of the trick requires some dexterity, and wrist movements are performed during the procedure. The mobile application core used for this research has been previously developed by a gaming company
using the Unity3D game engine, and runs on Android phones~\cite{alberts2022designing}.

In this study, we implement a movement recognition system to accurately recognize dorsiflexion movements during the use of this mobile application, as well as an intuitive adaptive difficulty system. Since the patients need to perform specific movements as part of their hand therapy, the accuracy of these movements needs to be assessed to evaluate therapy progression. Also, different patients have different needs and might not have the same range of motion in their wrists as other patients. Consequently, it is necessary that the application should adjust to the user, and keep adjusting if the capabilities of the user change over time. 

The remainder of the paper is structured as follows. First, we discuss related research in Section~\ref{section:related}. Next, we present our data collection approach for the movement classification system in Section~\ref{section:data}. In Section~\ref{section:classification} we describe the features we use, and the approaches tested for classification of dorsiflexion. The experimental results are summarized in Section~\ref{section:experiments}. In Section~\ref{section:adapting}, we present the adaptive difficulty system introduced to the application. Section~\ref{section:target} reports results obtained with the target patient group. Finally, we provide a discussion and conclusions in Section~\ref{section:conclusion}.


\section{Related work}
\label{section:related}
The potential of personal informatics (PI) systems for people with motor disabilities has been discussed in~\cite{motahar2022review}. Here we focus on games that target children with motor skill impairments specifically, as well as provide some references for mobile phone based activity recognition, which is a very broad area due to the potential richness of the activities that can be detected. This paper, however, is the first for detecting dorsiflexion movements in the literature. 
\subsection{Games targeted at children with motor skill impairments}
While many applied games in the field of health have been developed in the past, the selection of applied games that specifically target children with motor skill impairments is not as substantial~\cite{ayed2019vision, page2017active}. In~\cite{gurbuzsel2022eliciting}, a number of smart toys and game applications are listed to monitor and enhance fine motor skills of children, but these are not geared towards rehabilitation.
Researchers from the University of Amsterdam and the Amsterdam University of Applied Sciences developed a game for detecting delays in motor skill development on a commercially available toy called the \textit{Futurocube}~\cite{sander2017detecting}.
Using this game, they were able to predict the degree of motor skill impairment in the participants with increasing accuracy as the difficulty of the game rose.
More recent work with the Futurocube has focused on using different machine learning techniques for classifying children with motor skill impairments~\cite{brons2021assessing}.

\cite{mironcika2018smart} developed a smart toy for motor skill assessment in form of a board game. The goal of this game is to move tokens (depicting mice) across the board and turn them around as carefully as possible so as to not wake up `the cat'. The tokens contain accelerometers, which are used to assess the smoothness of movements by computing the mean squared jerk feature. The results show that children with better fine motor skills can move the tokens more smoothly using one hand, making the game suitable for detecting motor skill impairments. Moreover, the game was perceived as fun and exciting to play by the children participating in the experiment. More recently,~\cite{moreno2021utilization} presented a mobile application for analyzing the fine motor performance in children based on a similar idea, by making them move an object along several paths, varying in difficulty from straight to zigzag-shaped.

A smart toy that was developed for practicing fine motor skills in~\cite{serpa2021intelligent} encompassed several different exercises to practice different types of movements: tracing a path while holding a pen, placing clothespins on colored fabric, and carrying a metal ring from left to right while avoiding touching the filament, respectively. In most of these applications, the difficulty of the game is adjusted by having multiple challenges with different difficulty levels. In this paper, however, we measure and adapt the difficulty to the child. Measuring performance has the double purpose of observing rehabilitation progress longitudinally. 

\subsection{Human activity recognition using smartphone sensors}
A modern smartphone contains a variety of sensors, including an accelerometer, a magnetometer, and a gyroscope, which can be highly useful for recognizing the activities of a user~\cite{peng2018aroma,ravi2005activity, shoaib2014fusion}. Previous research has largely focused on human movement recognition using the sensors available in smartphones~\cite{pires2016data, micucci2017unimib, nweke2018deep, nweke2019data}. The most common application is the detection of everyday activities including sitting, walking, and walking up the stairs. For these types of applications, the phone is usually passively carried around. Some of these applications addressed disabilities or pathologies impairing movements of the limbs, such as gait~\cite{zhang2019pdmove}.

Deep learning methods have become widely used for activity recognition in recent years~\cite{bozkurt2021comparative}. Some examples of different types of networks previously used for activity recognition include basic multilayer perceptrons (MLPs)~\cite{wang2019deep}, convolutional neural networks (CNNs)~\cite{ignatov2018real}, and recurrent neural networks (RNNs)~\cite{murad2017deep}.

CNNs are mostly used for visual data, such as images and video, but have also been used on time series data such as sensor data collected using smartphone sensors~\cite{ignatov2018real, wang2019deep}. 
One study proposed the use of CNNs for activity recognition using the accelerometer in a mobile phone~\cite{ignatov2018real}. As input for the network, they used both the raw accelerometer data and a selection of extracted basic statistical features that describe the global properties of the time-series data. The combination of both seemed to achieve the best performance. For mobile phone applications, shallow architectures and lower computational costs are desirable, as battery life is an important consideration.

For supervised activity recognition tasks, the temporal dimension is important, as a large part of the challenge is to properly segment the activity boundaries and to provide continuous class labels. For this purpose, RNNs~\cite{murad2017deep} and long short-term memory (LSTM) networks were used~\cite{chen2016lstm}. 
LSTM layers can also be combined with convolutional layers to create more complex models for human activity recognition using raw mobile phone sensor data~\cite{xia2020lstm}. 

In our application, we examine a range of simple to complex models to classify dorsiflexion movements. What is more important than the selection of the machine learning approach is the insights into how a cerebral palsy patient would handle a mobile phone during the exergame, and how that will affect the performance of the classifiers. We first deal with the problem of data collection, as most ML approaches need as much data as possible to provide good classification results.


\section{Data collection}
\label{section:data}
To train a model for distinguishing dorsiflexion from other types of hand movements, we collected data for a variety of movements, as well as dorsiflexion movements from different starting positions. 
We have defined 28 distinct movement classes: the first 10 classes describe dorsiflexion movements from different starting positions, and the other 18 classes describe other kinds of movements and non-movements which function as the true negatives in the dataset, including rotations, shaking, and the phone sitting still.
We use the accelerometer and the gyroscope to record the data along three perpendicular axes ($x$, $y$, $z$). The phone that was used for data collection, and all subsequent training and evaluation, was a Samsung Galaxy A52.

We have created a custom mobile application for data collection, where the user is shown an animation showing how to perform a movement, along with auditory instructions and a text description of the movement. This interface selects ten movement classes per session randomly for each user. Additionally, we recorded the hand movements of the user using an external video camera, for improving annotation quality. A total of 37 data collection sessions with 20 unique subjects were completed. Before participating in the experiment, the subjects were informed of the purpose and methods of the study and signed a written consent form. The data collection and storage procedures were reviewed and approved by a medical ethics review committee\footnote{Details left out for blind review. The authors declare that there was no influence or involvement from the funding organization in the design, data collection, analysis, interpretation, writing, or submission of this study.}.

Two annotators manually annotated the beginning and end times of the movements using the annotation tool ELAN~\cite{elansoftware}.
Only a binary label was introduced (i.e. dorsiflexion vs. no dorsiflexion) during the annotations. 
The resulting dataset consists of a total number of 337 segments, out of which 113 are classified as dorsiflexion movements. 
The dataset is divided into a training (15 subjects) and test (5 subjects) set for model training and evaluation purposes. The test set contains 53 segments, representing 15.7\% of the total dataset.

\begin{figure}[thb!]
\centering
\includegraphics[width=0.8\textwidth]{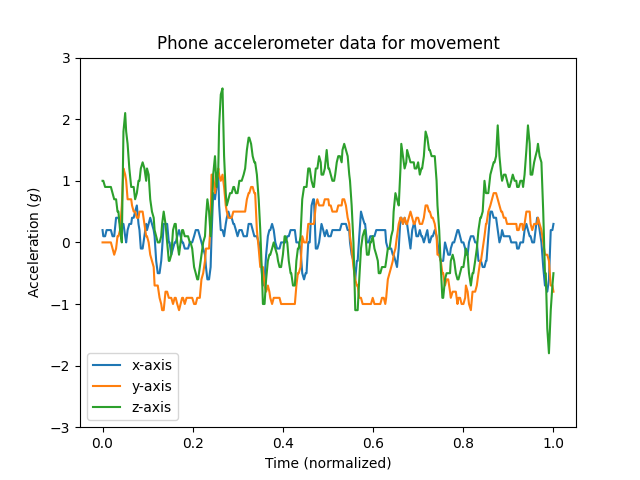}
\includegraphics[width=0.8\textwidth]{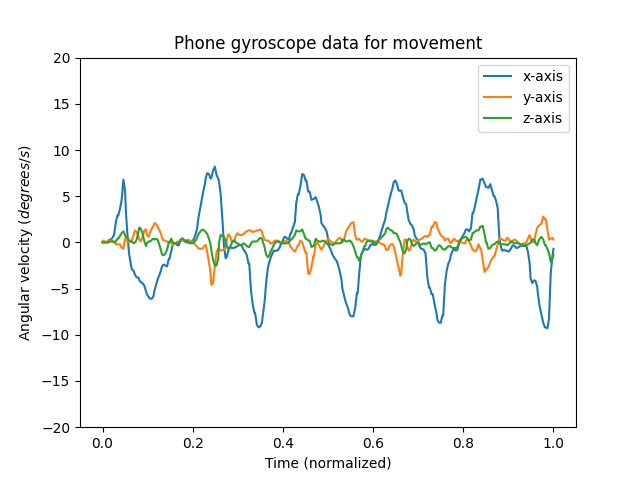}
\caption{Recorded sensor data for movement class 1: dorsiflexion with the phone upright and the screen turned away from the hand palm.}
\label{fig:sensordata1}
\end{figure}

\begin{figure}[thb!]
\centering
\includegraphics[width=0.8\textwidth]{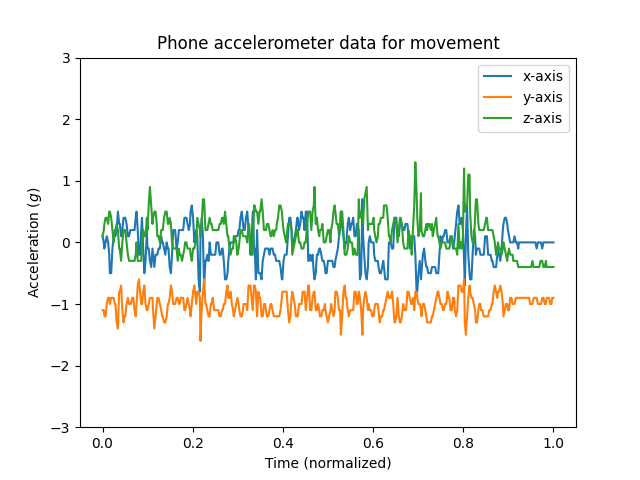}
\includegraphics[width=0.8\textwidth]{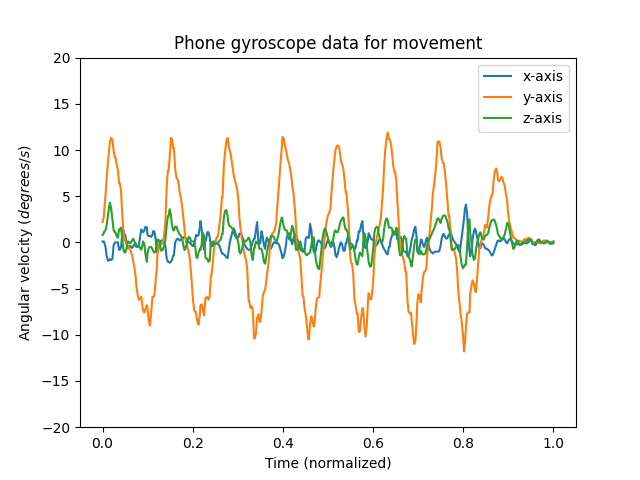}
\caption{Recorded sensor data for movement class 21: rotating the phone to the left and right side alternately, with the phone held horizontally and the screen turned upward.}
\label{fig:sensordata21}
\end{figure}

Figures~\ref{fig:sensordata1} and~\ref{fig:sensordata21} show two examples of what the data looks like, for one dorsiflexion movement and one non-dorsiflexion movement (i.e. rotating the phone), respectively. The collected accelerometer and gyroscope data are plotted for each of the three axes. From these plots, it can be seen that the recorded sensor data for these movements look similar at first sight, especially when looking at the gyroscope data.

\section{Classification of Dorsiflexion}
\label{section:classification}

\subsection{Feature extraction and selection}
We investigated both feature extraction and feature selection to train the machine learning models. For the former, we extracted seven low level descriptors (i.e. mean, min, max, standard deviation, variance, skew, and kurtosis, respectively) from each of the axes of the sensors, resulting in $6\times7=42$ features. Each feature was normalized using min-max normalization. 

For feature selection, we used the Minimum Redundancy Maximum Relevance (mRMR) algorithm~\cite{ding2005minimum}. This method, attempts to find an optimal number of features by eliminating redundant features. For each feature, a score is computed based on the maximum relevance with regard to the output variable, and minimal redundancy with regard to the already selected features. The algorithm iterates until $k$ features have been selected.
One advantage of mRMR is the explainability of the selected features, which is important because we want to be able to explain the logic behind the selected thresholds for the dorsiflexion movements.

There are various options for score functions to use with mRMR, and we used a function that combines the F-test statistic and the Pearson correlation coefficient to compute a score for each feature~\cite{zhao2019maximum}. This method is simple and fast, yet provides comparable results to other implementations of mRMR. The relevance of a feature is computed as the F-statistic between the feature and the target output. The redundancy of the feature is computed as the average Pearson correlation between the feature and the features selected during previous iterations. 

\subsection{Classification}

We tested and compared several machine learning approaches of different complexity for solving the classification problem. Since the mobile phone presents a limited energy budget, simple approaches are preferred to computation-heavy approaches. We tested k-nearest neighbors (k-NN), support vector machines (SVM), multi-layer perceptrons (MLP), convolutional neural networks (CNN), long-short-term-memory (LSTM), and Bidirectional LSTM, respectively. To evaluate the results, we used classification accuracy, precision, recall, and the F-score. 

A basic k-NN classifier is used as a baseline algorithm with the 42 features obtained via feature extraction, as described earlier. We used the Euclidean distance as the distance function, and leave-one-out cross-validation on the training partition was used to select the value for $k$ that provided the highest accuracy, which was set as 1. 

Next, we performed mRMR for determining the features to be used for the final model. The highest training set accuracy was observed when using 21 features out of 42, ultimately resulting in an accuracy of 0.997 on the training set for $k=1$ and 21 features. Therefore, these are the values that are used for evaluation on the test set. We have tested dynamic time warping (DTW) in conjunction with kNN, but DTW is too time-intensive to use in a real-time application, and was not taken into further consideration.


The second model we trained is a support vector machine (SVM) with a linear kernel, using the same features that were used with 1-NN. Using mRMR on the training set resulted in the selection of only 2 features out of the 42, with a cross-validation accuracy of 0.993. We also trained four neural network (NN) models. 


\subsubsection{Multilayer perceptron (MLP)}
The first NN we have trained is a multilayer perceptron, with two hidden layers (with 128 and 256 neurons, respectively), a dropout layer, and a classification layer. This network takes as input the same features that are also used for the k-NN and the SVM. We first used mRMR to select the appropriate number of features, computing the accuracy using leave-one-out cross-validation on the training set for each possible number of features. The network was trained with the ADAM optimizer for 100 epochs max~\cite{kingma2014adam}.
The highest scoring number of features was 37, with an accuracy of 0.958.


\subsubsection{Convolutional neural network (CNN)}
While CNNs are most commonly used for image recognition and classification tasks, they can also be applied to activity recognition tasks using sensor data. The network structure we used was based on a CNN used previously for activity recognition~\cite{ignatov2018real}. The first layer is a 1-dimensional convolutional layer with 196 filters. The kernel size used is 1 $\times$ 16. This convolutional layer is followed by a max pooling layer with pool size 4 and a flattening layer. Next is a dense layer with 1024 neurons, followed by the classification layer.

\subsubsection{Long Short-Term Memory}
We tested an RNN with LSTM layers~\cite{hochreiter1997long}. LSTMs have been shown to benefit from at least two or more LSTM layers, which is why our model had two LSTM layers, with 128 and 256 nodes respectively. They are followed by a dropout layer and the classification layer.

\subsubsection{Bidirectional Long Short-Term Memory}
The BLSTM was constructed similarly to the LSTM, but instead of two LSTM layers, it contains two BLSTM layers, also with 128 and 256 nodes respectively~\cite{schuster1997bidirectional}. Similar to the LSTM, the BLSTM layers are followed by a dropout layer and the classification layer.

\section{Experimental Results}
\label{section:experiments}
The experimental results of each trained model can be seen in Table~\ref{table:dorsiflexionresults}. The accuracy, precision, recall, and F-score are shown for each of the models, reported on the independent test set. The precision, recall, and F-score are also shown for each class separately. The highest scores are shown in bold text for clarity.

The CNN model performed well overall, but the precision and recall results indicate that most of the time dorsiflexion movements are indeed classified as such, but there are also false positives. This is made evident by the fact that the F-score for these models is lower. The more complex NN models show more balanced scores and higher F-scores, with CNN scoring the highest overall, suggesting that it is a suitable machine learning model for classifying dorsiflexion in a binary classification task.

\begin{table}[h]
\centering
\setlength{\tabcolsep}{4pt}
\renewcommand{\arraystretch}{1.05}
\begin{tabular}{llllllll}
\multicolumn{1}{l}{}                                           & \multicolumn{1}{l}{\textbf{}}          & \multicolumn{1}{l}{\textbf{KNN*}} & \multicolumn{1}{l}{\textbf{SVM*}} & \multicolumn{1}{l}{\textbf{MLP*}} & \multicolumn{1}{l}{\textbf{CNN}} & \multicolumn{1}{l}{\textbf{LSTM}} & \multicolumn{1}{l}{\textbf{BLSTM}} \\ 
\toprule
\multicolumn{1}{l}{\multirow{4}{*}{\textbf{Overall}}}          & \multicolumn{1}{l}{\textbf{Accuracy}}  & \multicolumn{1}{l}{0.943}         & \multicolumn{1}{l}{0.925}         & \multicolumn{1}{l}{0.962}              & \multicolumn{1}{l}{\textbf{0.970}}        & \multicolumn{1}{l}{0.961}         & \multicolumn{1}{l}{0.964}            \\ 
\multicolumn{1}{l}{}                                           & \multicolumn{1}{l}{\textbf{Precision}} & \multicolumn{1}{l}{0.932}         & \multicolumn{1}{l}{0.913}         & \multicolumn{1}{l}{0.952}              & \multicolumn{1}{l}{\textbf{0.963}}        & \multicolumn{1}{l}{0.949}         & \multicolumn{1}{l}{0.953}            \\ 
\multicolumn{1}{l}{}                                           & \multicolumn{1}{l}{\textbf{Recall}}    & \multicolumn{1}{l}{0.956}         & \multicolumn{1}{l}{0.941}         & \multicolumn{1}{l}{\textbf{0.971}}              & \multicolumn{1}{l}{\textbf{0.971}}        & \multicolumn{1}{l}{0.968}         & \multicolumn{1}{l}{\textbf{0.971}}            \\ 
\multicolumn{1}{l}{}                                           & \multicolumn{1}{l}{\textbf{F-score}}  & \multicolumn{1}{l}{0.940}         & \multicolumn{1}{l}{0.921}         & \multicolumn{1}{l}{0.960}              & \multicolumn{1}{l}{\textbf{0.967}}        & \multicolumn{1}{l}{0.957}         & \multicolumn{1}{l}{0.961}            \\ \midrule
\multicolumn{1}{l}{\multirow{3}{*}{\textbf{Dorsiflexion}}}     & \multicolumn{1}{l}{\textbf{Precision}} & \multicolumn{1}{l}{0.864}         & \multicolumn{1}{l}{0.826}         & \multicolumn{1}{l}{0.905}              & \multicolumn{1}{l}{\textbf{0.938}}        & \multicolumn{1}{l}{0.903}         & \multicolumn{1}{l}{0.911}            \\ 
\multicolumn{1}{l}{}                                           & \multicolumn{1}{l}{\textbf{Recall}}    & \multicolumn{1}{l}{\textbf{1.000}}         & \multicolumn{1}{l}{\textbf{1.000}}         & \multicolumn{1}{l}{\textbf{1.000}}              & \multicolumn{1}{l}{0.974}        & \multicolumn{1}{l}{0.991}         & \multicolumn{1}{l}{0.991}            \\ 
\multicolumn{1}{l}{}                                           & \multicolumn{1}{l}{\textbf{F-score}}  & \multicolumn{1}{l}{0.927}         & \multicolumn{1}{l}{0.905}         & \multicolumn{1}{l}{0.950}              & \multicolumn{1}{l}{\textbf{0.956}}        & \multicolumn{1}{l}{0.945}         & \multicolumn{1}{l}{0.949}            \\ \midrule
\multicolumn{1}{l}{\multirow{3}{*}{\textbf{Non-dorsiflexion}}} & \multicolumn{1}{l}{\textbf{Precision}} & \multicolumn{1}{l}{\textbf{1.000}}         & \multicolumn{1}{l}{\textbf{1.000}}         & \multicolumn{1}{l}{\textbf{1.000}}              & \multicolumn{1}{l}{0.987}        & \multicolumn{1}{l}{0.995}         & \multicolumn{1}{l}{0.995}            \\ 
\multicolumn{1}{l}{}                                           & \multicolumn{1}{l}{\textbf{Recall}}    & \multicolumn{1}{l}{0.912}         & \multicolumn{1}{l}{0.882}         & \multicolumn{1}{l}{0.941}              & \multicolumn{1}{l}{\textbf{0.968}}        & \multicolumn{1}{l}{0.945}         & \multicolumn{1}{l}{0.950}            \\ 
\multicolumn{1}{l}{}                                           & \multicolumn{1}{l}{\textbf{F-score}}  & \multicolumn{1}{l}{0.954}         & \multicolumn{1}{l}{0.938}         & \multicolumn{1}{l}{0.970}              & \multicolumn{1}{l}{\textbf{0.977}}        & \multicolumn{1}{l}{0.970}         & \multicolumn{1}{l}{0.972}            \\ \bottomrule
\multicolumn{8}{l}{* uses extracted features as input, as opposed to raw sensor data}  
\end{tabular}
\caption{The accuracy, precision, recall, and F-score for dorsiflexion recognition per model on the test set.}
\label{table:dorsiflexionresults}
\end{table}


\section{Adapting the Game to the User}
\label{section:adapting}
When a movement has been positively identified as a dorsiflexion movement, we want to check at what level the user is shaking the phone, using an adaptive difficulty system. The mobile application is supplied with two indicators for this purpose, namely, the range of motion and speed. To achieve adaptive difficulty, the application needs to be able to adjust the sensitivity to the user with regards to these two measures, and slowly decrease the sensitivity to make it harder to get the correct movement. This feature helps to get the user to progress in their therapy and improve their motor skills.

\subsection{Rule-based adaptive difficulty}
We implemented two decision rules to determine at which points the thresholds for both the range of motion and the speed should change. To determine when to change the difficulty, the past performance of the player and the percentage of correct movements over the last 10 registered shakes is analysed.
If at least 90\% of the movements reached the threshold, the threshold increases; if no more than 60\% of the movements remained under the threshold, the threshold decreases. 

\subsection{Calibration}
Since we wanted to adjust the dorsiflexion recognition sensitivity to the user, the application first goes through a calibration phase when it is used for the first time. During this step, the player performs dorsiflexion to activate the different stages of the `magic trick', and the adaptive system uses the first five movements to determine the thresholds for both the range of motion and the speed.
Besides the automatic calibration, the user can also reset the thresholds and restart the calibration from the settings screen, or adjust the thresholds manually if they wish. This is a precaution against the application not functioning according to the preferences of the user, who may prefer easier or more challenging settings.

\subsection{Range of motion}
In order to train the system, a pilot was conducted without accessing the target patient group. 

\begin{figure}[h]
\centering
\includegraphics[height=6cm]{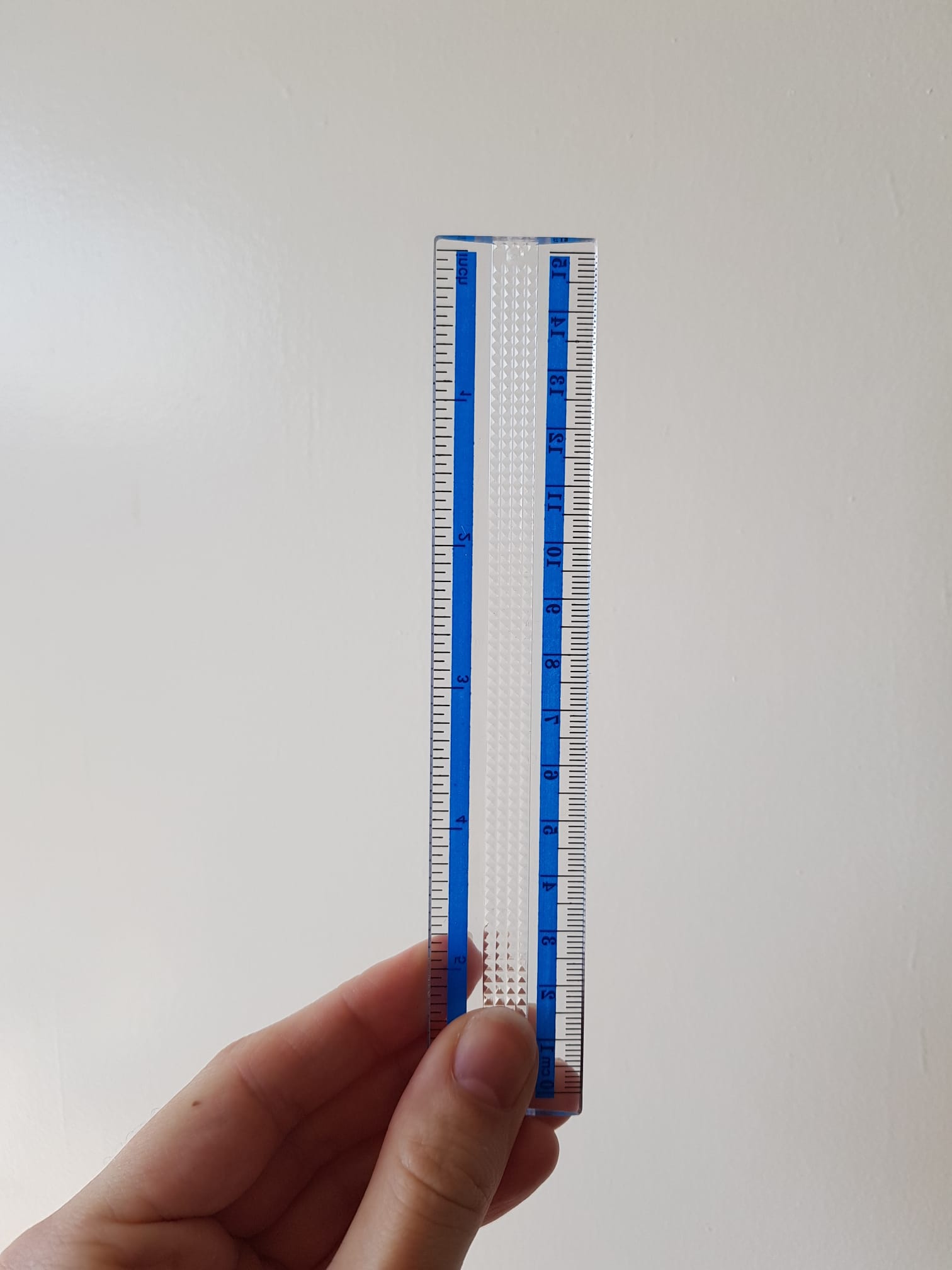}
\includegraphics[height=6cm]{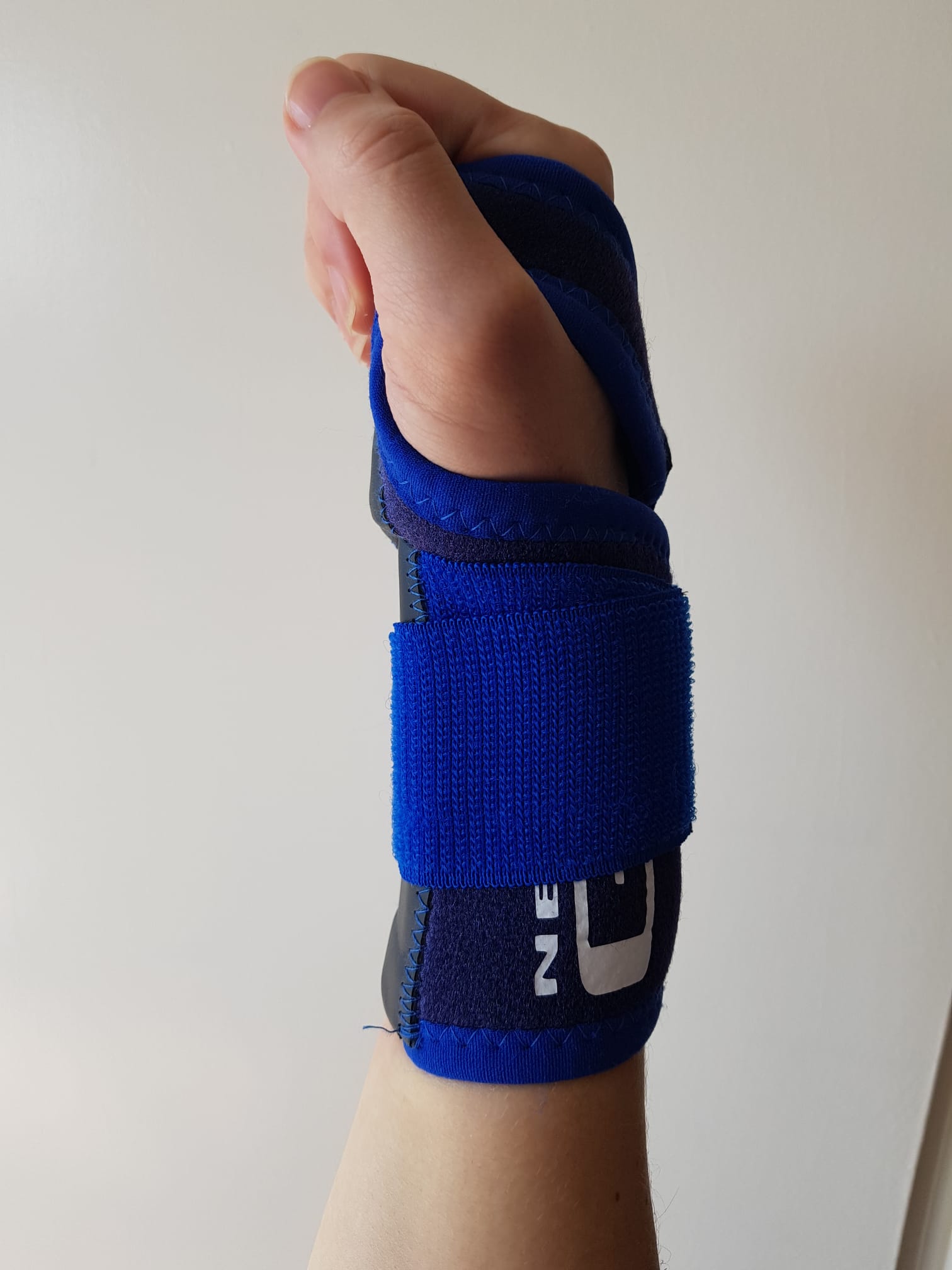}
\includegraphics[height=6cm]{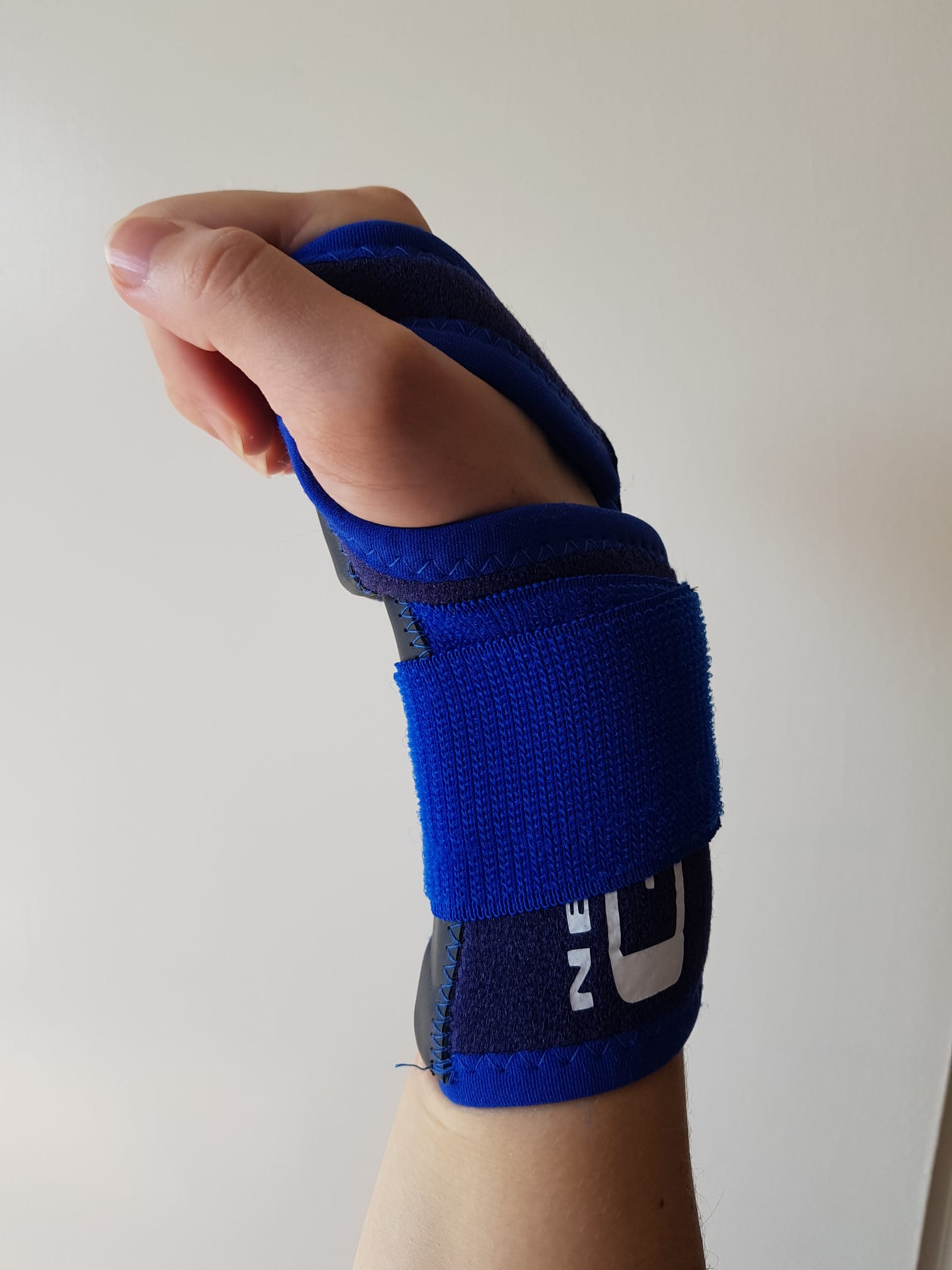}
\caption{Dorsiflexion while wearing an arm brace with a plastic ruler inserted.}
\label{fig:halfrestricted}
\end{figure}

Children with hemiparesis as a result of cerebral palsy often wear an arm brace to support their wrist, such as the one shown in Figure~\ref{fig:halfrestricted}. However, besides supporting the wrist, this arm brace also restricts the movement of the wrist. Subsequently, we used arm braces to simulate restricted wrist movements for different ranges of motion.

\begin{figure}[h]
\centering
\includegraphics[height=6cm]{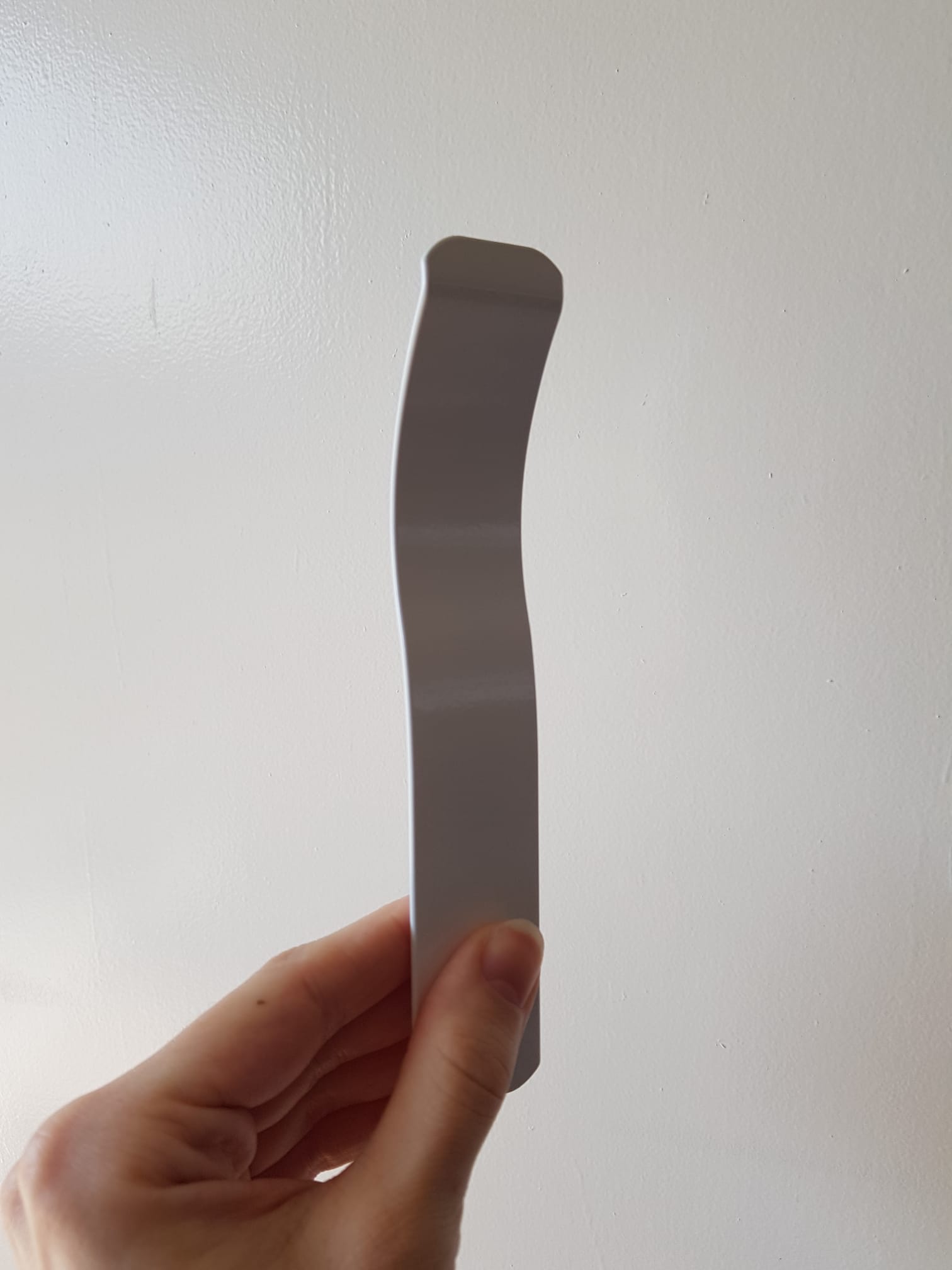}
\includegraphics[height=6cm]{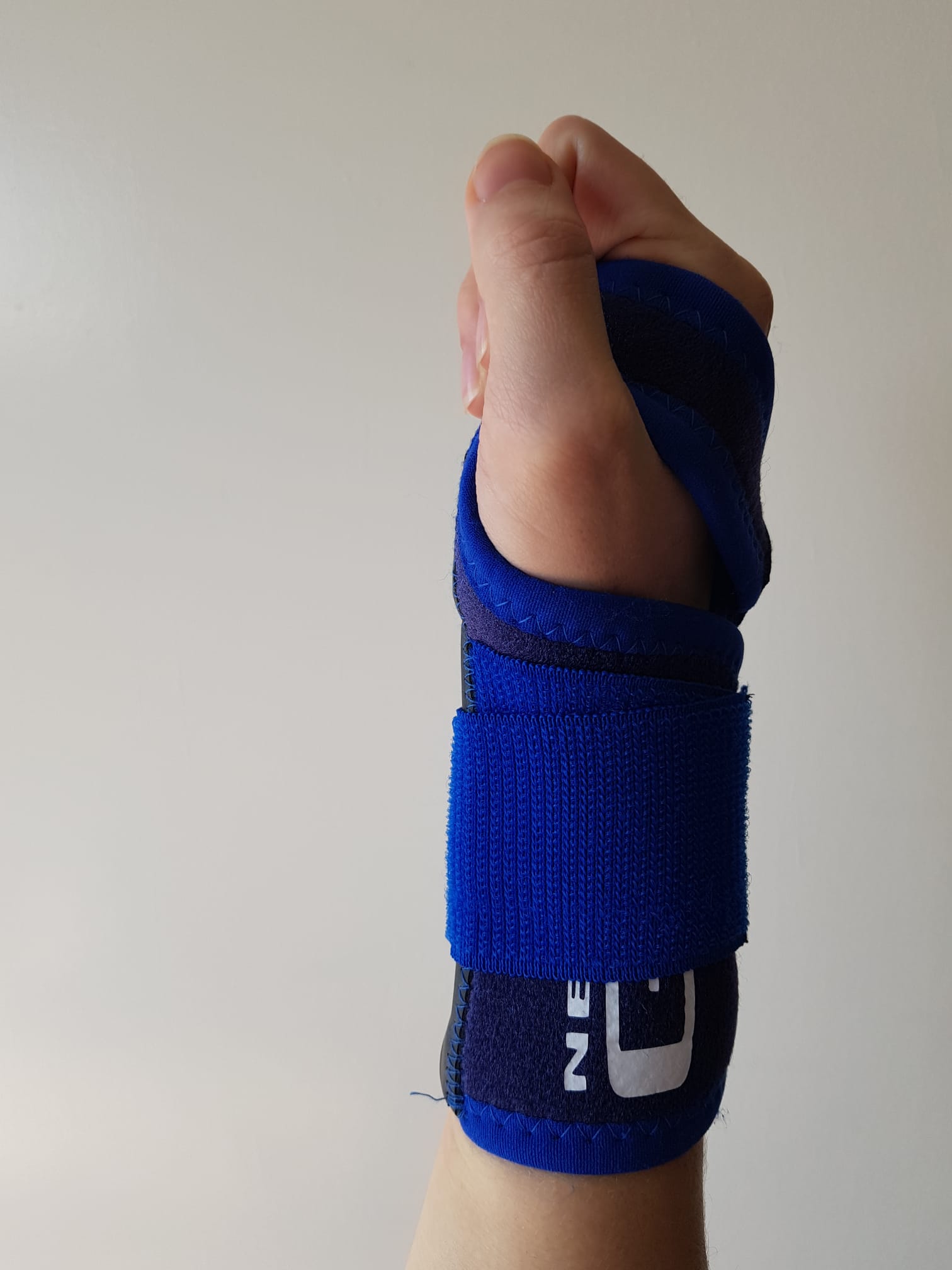}
\includegraphics[height=6cm]{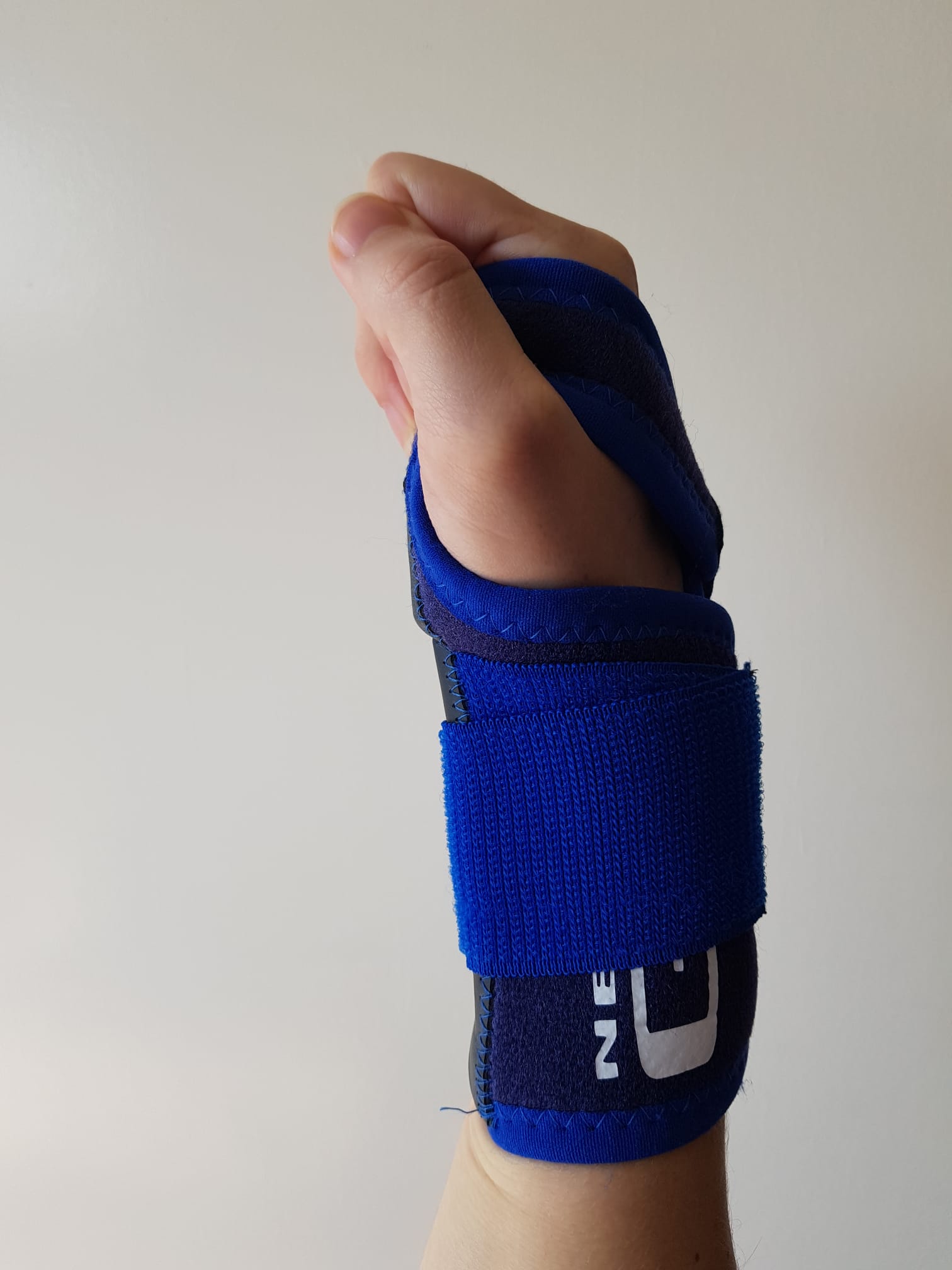}
\caption{Dorsiflexion while wearing an arm brace with a metal bar ruler inserted.}
\label{fig:fullyrestricted}
\end{figure}

Data were collected for the following conditions of dorsiflexion, all from the same starting position to make comparing the movements easier:
\begin{itemize}
\item Free movement without the arm brace
\item Half-restricted movement with a plastic ruler inserted in the arm brace, resulting in the range of motion seen in Figure~\ref{fig:halfrestricted}
\item Fully-restricted movement with a metal ruler inserted in the arm brace, resulting in the range of motion seen in Figure~\ref{fig:fullyrestricted}
\end{itemize}
These three degrees of movement are considered the three different movement classes. 
For each of these movements, we recorded 20 samples. We performed feature extraction and selection as with the previously described experiments. The idea is to obtain features that are suitable to recognize varying strengths of dorsiflexion, which can be used to set a threshold to determine the range of motion of the dorsiflexion movement. 

\begin{figure}[thb!]
\centering
\includegraphics[width = 0.8\textwidth]{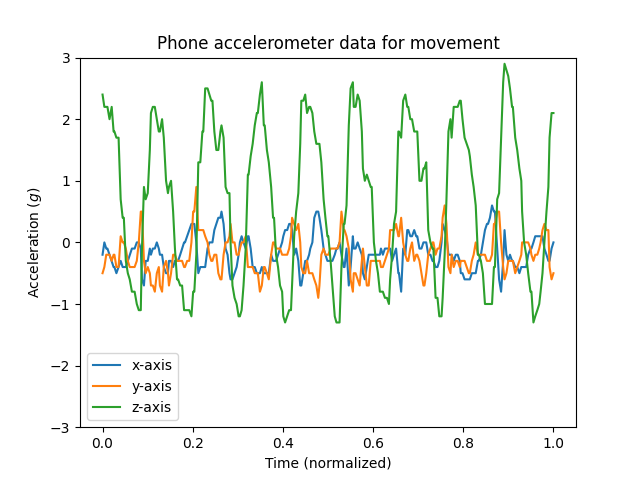}
\includegraphics[width = 0.8\textwidth]{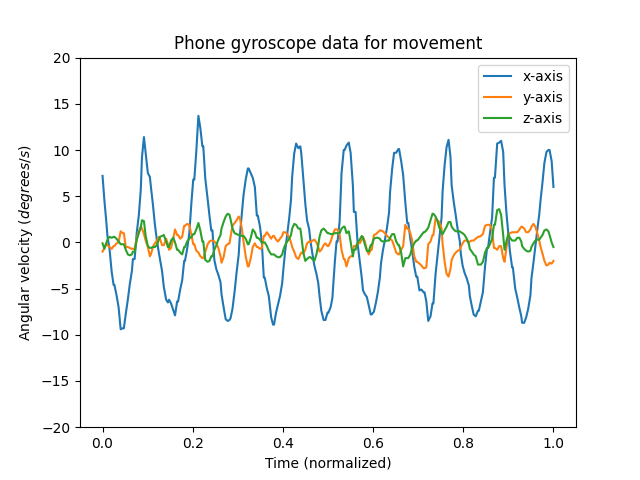}
\caption{Sensor data for dorsiflexion, with the wrist unrestricted.}
\label{fig:rangeofmotionunrestricted}
\end{figure}

\begin{figure}[thb!]
\centering
\includegraphics[width = 0.8\textwidth]{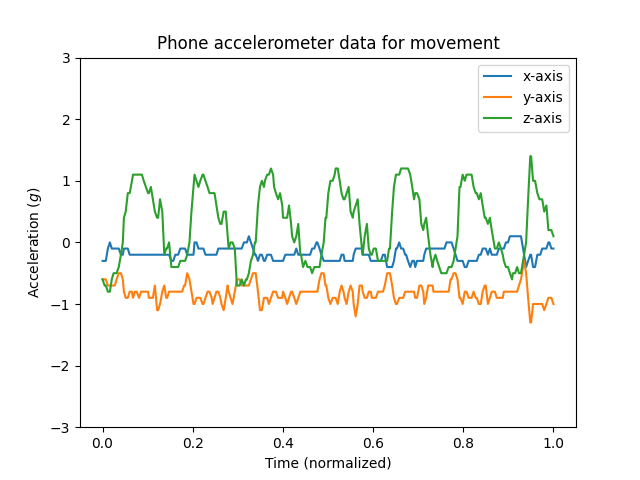}
\includegraphics[width = 0.8\textwidth]{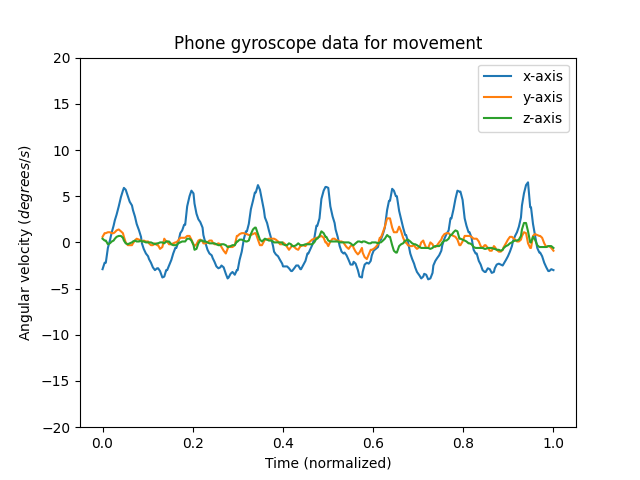}
\caption{Sensor data for dorsiflexion, while wearing the arm brace with a ruler inserted.}
\label{fig:rangeofmotionhalf}
\end{figure}

\begin{figure}[thb!]
\centering
\includegraphics[width = 0.8\textwidth]{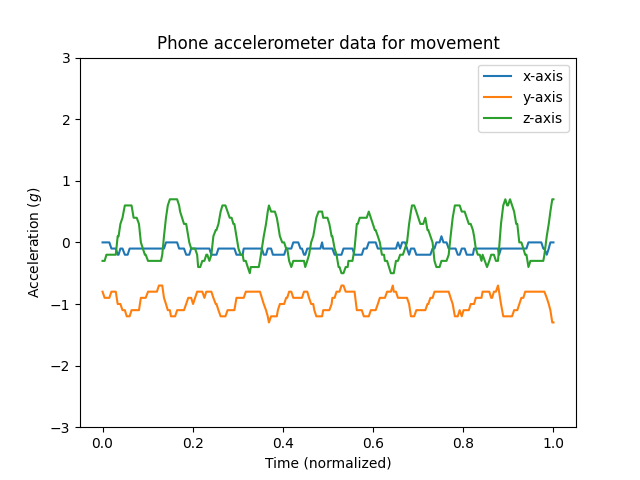}
\includegraphics[width = 0.8\textwidth]{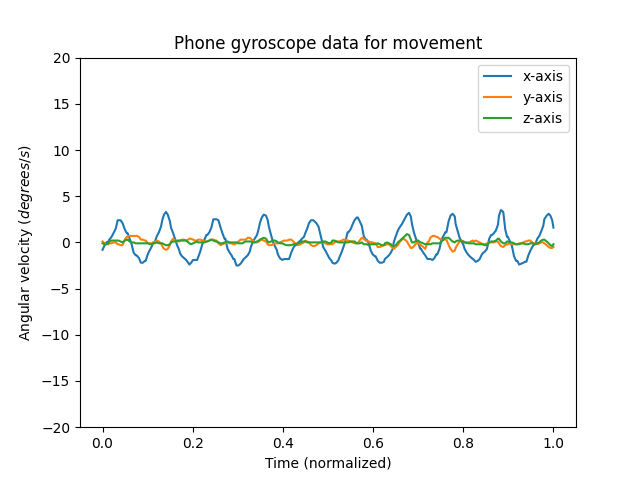}
\caption{Sensor data for dorsiflexion, while wearing the arm brace with a metal bar inserted.}
\label{fig:rangeofmotionfully}
\end{figure}

A visualization of some of the collected sensor data, for each level of restriction, can be seen in Figures~\ref{fig:rangeofmotionunrestricted},~\ref{fig:rangeofmotionhalf}, and~\ref{fig:rangeofmotionfully}.

We have considered the mean absolute error (MAE), mean squared error (MSE), and root mean squared error (RMSE) to measure the performance of regression models. RMSE is expressed in the same units as the original $y$ values, and penalizes higher error values more compared to MAE. Therefore, we have used RMSE for evaluating the models. 

For each set of features, leave-one-out cross-validation was used to compute the RMSE for that number of features. Using a single feature (namely, the standard deviation of the $x$-axis of the gyroscope) ultimately results in the lowest RMSE. On the test set, the model results in an RMSE of 1.419. Since a single feature is powerful enough to determine the range of motion for the most part, the adaptive difficulty system was linked to this indicator, together with the simple threshold adjustment rules described earlier. This provided a simple and reliable method for evaluating the range of motion of the user. 

\subsection{Speed}
\begin{figure}[thb!]
\centering
\includegraphics[width = 0.8\textwidth]{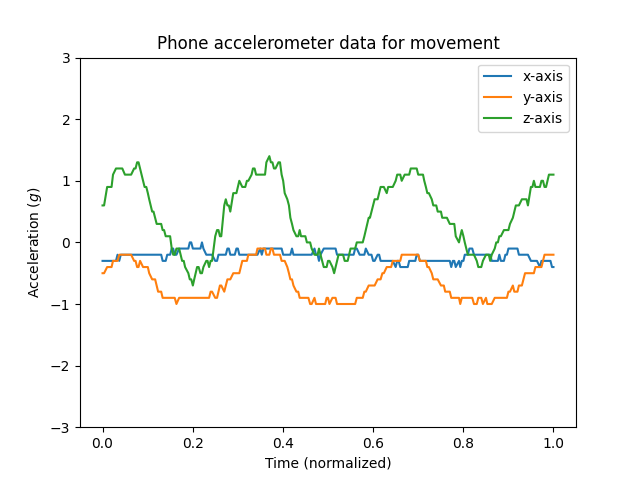}
\includegraphics[width =0.8 \textwidth]{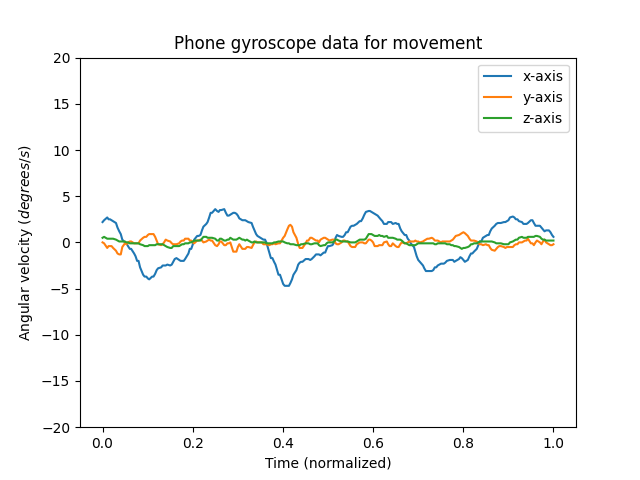}
\caption{Sensor data for dorsiflexion, performed with low speed.}
\label{fig:fastnessslow}
\end{figure}

\begin{figure}[thb!]
\centering
\includegraphics[width = 0.8\textwidth]{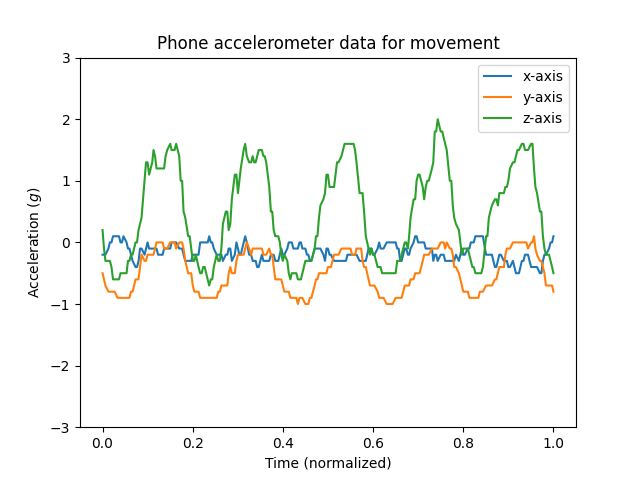}
\includegraphics[width = 0.8\textwidth]{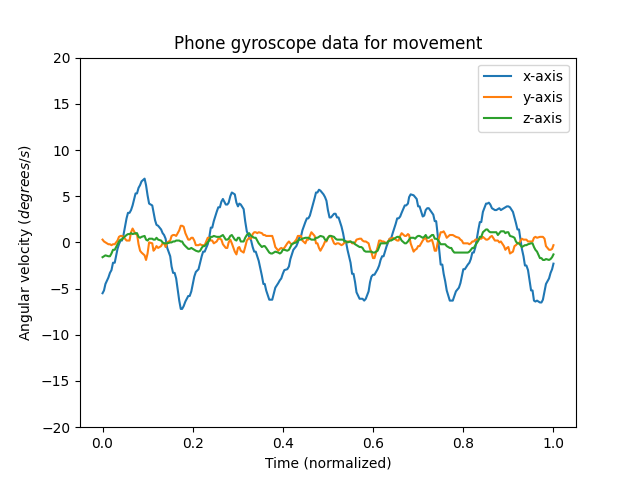}
\caption{Sensor data for dorsiflexion, performed at medium speed.}
\label{fig:fastnessmedium}
\end{figure}

\begin{figure}[thb!]
\centering
\includegraphics[width = 0.8\textwidth]{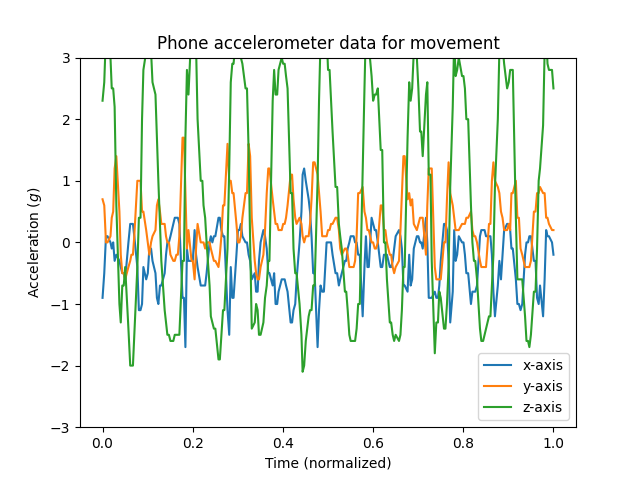}
\includegraphics[width = 0.8\textwidth]{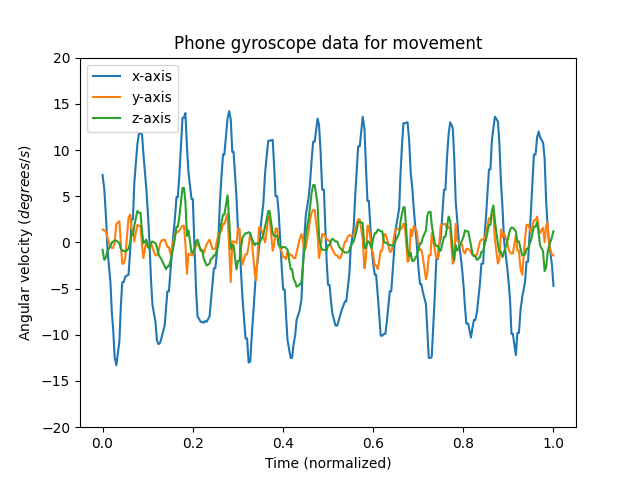}
\caption{Sensor data for dorsiflexion, performed at fast speed.}
\label{fig:fastnessfast}
\end{figure}

We collected data for dorsiflexion movements with the phone upright for three degrees of speed, classified as slow, medium, and fast, respectively. These classes were used to provide some qualitative insights, and examples of the collected sensor data can be seen in Figures~\ref{fig:fastnessslow},~\ref{fig:fastnessmedium}, and~\ref{fig:fastnessfast}. A clear difference can be observed between different movement speeds. Most notably, the maximum amplitude of both the accelerometer and gyroscope signals goes up as speed increases along the axis on which the movement occurred. As can be observed, the signals are similar to the different ranges of motion as seen in Figures~\ref{fig:rangeofmotionunrestricted},~\ref{fig:rangeofmotionhalf}, and~\ref{fig:rangeofmotionfully}. However, what sets the speed apart from the range of motion, is the frequency at which the minimum and maximum amplitudes alternate.

To compute the frequency, there are some different options, including:
\begin{itemize}
    \item Computing a fast Fourier transform (FFT) and taking the lowest frequency.
    \item Calculating the number of zero crossings or the number of times the sensor value oscillates about 0. The faster the sensor value oscillates around 0, the higher the frequency.
\end{itemize}

In our implementation, we used the number of zero crossings for the speed. While this is a rigid scale and does not allow for a gradual increase in difficulty, it does provide clear levels for the user to achieve, and it is intuitively interpreted.

\section{Tests with the Target Group}
\label{section:target}
Using the implemented dorsiflexion recognition and adaptive difficulty systems as described above, we performed an evaluation test with the target group to check the accuracy of the dorsiflexion recognition system. We collected data from eight children with cerebral palsy during therapy sessions at two clinics that specialize in treating posture and movement-related disorders. We traveled to the clinics and performed one-on-one sessions with each participant. Each participant was given a demonstration of how the magic monster is supposed to be used and was then asked to play with the magic monster application for three minutes. We obtained ethical approval from XXX prior to the study. The sessions were recorded using a video camera. The parents of the children gave written consent before the session with the magic monster, after informing them of the purpose and data collection procedure of the experiment.

One of the findings of this study, besides the data collection, was that the magic monster can sometimes be difficult and too big to hold for children with cerebral palsy. This is why a strap was added to the back of the monster, which wraps around the hand to hold the monster in place. This strap was added to the monster halfway through the testing phase, and could potentially have influenced the results.

Similarly to the previous data annotation that was done for training the machine learning algorithm, two people manually annotated the beginning and end times of the recorded movements in all of the recorded videos using ELAN~\cite{elansoftware}. The annotated dorsiflexion movements were then manually compared to the sensor data recorded using the phone. The mobile application checks if it detects dorsiflexion once per second to prevent performance issues by checking more often. The ground truth is obtained by inspection of the videos. The confusion matrix can be seen in Table~\ref{table:annotations}.

\begin{table}[h]
\begin{centering}
\begin{tabular}{llll}                                                                                       \\ \cline{3-4} 
    & \multicolumn{1}{l|}{}         & \multicolumn{2}{l|}{Annotated (real) values}                                                        \\ \cline{3-4} 
    & \multicolumn{1}{l|}{}         & \multicolumn{1}{l|}{Positive}                    & \multicolumn{1}{l|}{Negative}                    \\ \hline
\multicolumn{1}{|c|}{}                                  & \multicolumn{1}{l|}{Positive} & \multicolumn{1}{l|}{\cellcolor[HTML]{9AFF99}252} & \multicolumn{1}{l|}{\cellcolor[HTML]{FFCCC9}10}  \\ 
\multicolumn{1}{|c|}{\multirow{-2}{*}{Predicted by AI}} & \multicolumn{1}{l|}{Negative} & \multicolumn{1}{l|}{\cellcolor[HTML]{FFCCC9}35}  & \multicolumn{1}{l|}{\cellcolor[HTML]{9AFF99}931} \\ \hline
         &     &   &                                            
\end{tabular}

\caption{Confusion matrix for the binary dorsiflexion recognition.}
\label{table:annotations}
\end{centering}
\end{table}

To further evaluate the results, we computed the accuracy (0.948), precision (0.929), recall (0.846), and F-scores (0.885). 
Overall, the scores are high, with recall being lower than precision overall, meaning that not all dorsiflexion movements annotated as such are indeed being recognized as dorsiflexion. One limitation is that the target group is very small and it is difficult to obtain data from children with cerebral palsy. Our training set did not completely reflect the variation in the target group, and was collected from adult subjects. 

\section{Conclusions}
\label{section:conclusion}
In this study, we proposed a simple approach to detect dorsiflexion movements, and incorporated it into an application for hand and wrist rehabilitation of children with cerebral palsy. Our approach uses accelerometer and gyroscope data and uses a convolutional deep neural network on simple features extracted from these sensors to classify dorsiflexion movements.

Based on the classifier's output, the speed and range of motion were computed into simple indicators, which were added to the exergame to let it adapt to the user. We performed a series of experiments to measure the efficacy of the approach, and also tested the adaptive exergame on children with cerebral palsy.



The study has several shortcomings. More extensive data collection is needed for understanding the user experience aspects with the target group. This could be done with either the data collection application developed for this study, or with the magic monster application itself. While application data can be collected relatively easily, the video data that need to be collected for annotation and quality assessment purposes is much more difficult to obtain, and more sensitive to handle. However, once the annotations are completed, the video data can be destroyed, and only the anonymous gyroscope and accelerometer data can be shared with the annotations. 


The data collection was mostly focused on training an activity recognition model, and the adaptive difficulty adjustment of the system was developed with much less data. However, the logic of adaptation is fairly simple, and in general, less data will be sufficient for its development. Nonetheless, there is a need for a longitudinal user study, since the effects of having an adaptive difficulty can only be measured if the application is used for a longer period of time, and against a control group that uses a non-adaptive version of the application. Longitudinal studies can also assess the appeal of the application, and find out whether the novelty wears off, or whether the application loses its likeability after the magic trick is fully mastered.

The current data collection and evaluation were done using only one phone model. For future research, it would be fruitful to collect data using different models and compare the results, since the gyroscopes included in different phones might show slightly different behaviors. An additional shortcoming of the evaluation tests with the target group is that a strap was added to the back of the monster halfway through the data collection, potentially influencing the results.

We have not explored data augmentation strategies, but since data collection is both costly and time-consuming, data augmentation might help increase the number of available data points and increase the versatility of trained models. Data augmentation could potentially help increase the performance and reliability of the trained models when applying them to real-time applications.

Children with cerebral palsy stand to benefit from innovative solutions for aiding them in the arduous rehabilitation process. We believe the developed application can serve as a valuable tool, and the sub-modules can be integrated into other applications.  



\section*{Acknowledgments}
This study was funded by Utrecht University, the Amsterdam University of Applied Sciences, Eindhoven University of Technology 
as part of the Smart Technologies Empowering Citizens 
project, an in collaboration with 
Phillips and Ijsfontein and the HUMAN-AI fund. 
The authors declare that there was no influence or involvement from the funding organization in the design, data collection, analysis, interpretation, writing, or submission of this study.

\bibliographystyle{ACM-Reference-Format}
\bibliography{references}

\end{document}